\documentclass{article}
\usepackage{spconf,amsmath,graphicx,subfigure,float}
\usepackage{booktabs}
\usepackage{microtype}
\usepackage{flushend}
\usepackage{hyperref}
\usepackage{fancyhdr}
\usepackage{textcomp}
{
\fancyhf{}
\fancyfoot[R]{}}

\fancypagestyle{firstpage}{%
  \lfoot{\textcopyright 2019 IEEE.  Personal use of this material is permitted.  Permission from IEEE must be obtained for all other uses, in any current or future media, including reprinting/republishing this material for advertising or promotional purposes, creating new collective works, for resale or redistribution to servers or lists, or reuse of any copyrighted component of this work in other works.}
}

\title{Speech denoising by parametric resynthesis}
\name{Soumi Maiti and Michael I Mandel}
\address{The CUNY Graduate Center and Brooklyn College \\ New York, NY}
\begin{document}
%

\maketitle
\thispagestyle{firstpage}

\begin{abstract}
This work proposes the use of clean speech vocoder parameters as the target for a neural network performing speech enhancement. These parameters have been designed for text-to-speech synthesis so that they both produce high-quality resyntheses and also are straightforward to model with neural networks, but have not been utilized in speech enhancement until now.  In comparison to a matched text-to-speech system that is given the ground truth transcripts of the noisy speech, our model is able to produce more natural speech because it has access to the true prosody in the noisy speech. In comparison to two denoising systems, the oracle Wiener mask and a DNN-based mask predictor, our model equals the oracle Wiener mask in subjective quality and intelligibility and surpasses the realistic system.  A vocoder-based upper bound shows that there is still room for improvement with this approach beyond the oracle Wiener mask. We test speaker-dependence with two speakers and show that a single model can be used for multiple speakers.
\end{abstract}

\section{Introduction}
\label{sec:intro}
The general approach of speech enhancement has been to modify a noisy signal to make it more like the clean signal~\cite{WangSupervisedSpeechSeparation2018}. The main problems for such systems are the over-suppression of the speech and under-suppression of the noise. Ideally, speech enhancement systems should remove the noise completely without decreasing the speech quality. There are, however, statistical text-to-speech (TTS) synthesis systems that can produce high-quality speech from textual inputs (e.g., ~\cite{wu2016merlin}) by training an acoustic model to map text to the time-varying acoustic parameters of a vocoder, which then generates the speech. The most difficult part of this task, however, is predicting realistic prosody (timing information and pitch and loudness contours) from pure text. 

In this paper, we propose combining these two approaches to capitalize on the strengths of each by predicting the acoustic parameters of clean speech from a noisy observation and then using a vocoder to synthesize the speech.  We show that this combined system can produce high-quality and noise-free speech utilizing the true prosody observed in the noisy speech. We demonstrate that the noisy speech signal has more information about the clean speech than its transcript does. Specifically, it is easier to predict realistic prosody from the noisy speech than from text. Thus, we train a neural network to learn the mapping from noisy speech features to the acoustic parameters of the corresponding clean speech. From the predicted acoustic features, we generate clean speech using a speech synthesis vocoder. Since we are creating a clean resynthesis of the noisy signal, the output speech quality will be higher than standard speech denoising systems and completely noise-free. We refer to the proposed model as \emph{parametric resynthesis}.

In this paper, we show that parametric resynthesis outperforms statistical TTS in terms of traditional speech synthesis objective metrics.  Next we subjectively evaluate the intelligibility and quality of the resynthesized speech and compare it with a mask predicted by a DNN-based system~\cite{WangTrainingTargetsSupervised2014} and the oracle Wiener mask~\cite{ErdoganPhasesensitiverecognitionboostedspeech2015}. We show that the resynthesized speech is noise-free and has overall quality and intelligibility equivalent to the oracle Wiener mask and exceeding that of the DNN-predicted mask. We also show that a single parametric resynthesis model can be used for multiple speakers.

\section{Related Work}
\label{sec:back}
Traditional speech synthesis systems are of two types, concatenative and parametric. In our previous works, \cite{mandel14c, maiti2017concatenative, maiti2018large, syed2018concatenative} we proposed concatenative synthesis systems for denoising speech. Though these models can generate high quality speech, they are speaker-dependent and generally require a large dictionary of speech examples from that speaker. Alternatively, the current paper utilizes a parametric speech synthesis model, which more easily generalizes to combinations of conditions not seen explicitly in training examples.

In terms of parametric resynthesis, Rethage et al.~\cite{rethage2018wavenet} built an end-to-end model to map noisy audio to explicit models of both clean speech and noise using a WaveNet-like~\cite{van2016wavenet} architecture.  Compared to this model, our denoising system is much simpler, as it does not require an explicit model of the observed noise in order to converge and needs much less data and time to train.  This simplicity comes from using the non-neural WORLD vocoder~\cite{morise2016world}.



\begin{figure}[htb]
\begin{minipage}[b]{1.0\linewidth}
  \centering
  \includegraphics[trim={0.15cm 0 0.35cm 0},clip, width=1.0\textwidth,height=1.4cm]{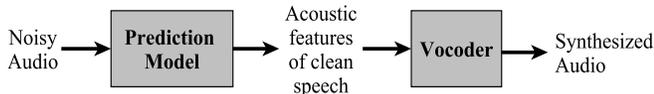}
\end{minipage}
\caption{Vocoder denoising model}
\label{fig:model}
\end{figure}

\section{Model overview }
\label{sec:mod_ovr}
Parametric resynthesis consists of two stages: prediction and synthesis as shown in Figure~\ref{fig:model}. The first stage is to train a prediction model with noisy audio features as input and clean acoustic features as output labels. The second stage is to resynthesize audio using the vocoder from the predicted acoustic features.

We use the WORLD vocoder~\cite{morise2016world} to transform between acoustic parameters and clean speech waveform. This vocoder allows both the encoding of speech audio into acoustic parameters and the decoding of acoustic parameters back into audio with very little loss of speech quality.  The acoustic parameters are much easier to predict using neural network prediction models than the raw audio.  We use the encoding of clean speech to generate our training targets and the decoding of predictions to generate output audio.  The WORLD vocoder is incorporated into the Merlin neural network-based speech synthesis system~\cite{wu2016merlin}, and we utilize Merlin's training targets and losses for our model.




The prediction model is a neural network that takes as input the log mel spectra of the noisy audio and predicts clean speech acoustic features at a fixed frame rate. The WORLD encoder outputs four acoustic parameters: i) spectral envelope, ii) log fundamental frequency (F0), iii) a voiced/unvoiced decision and iv) aperiodic energy of the spectral envelope. All the features are concatenated with their first and second derivatives and used as the targets of the prediction model. There are 60 features for spectral envelope, 5 for band aperiodicity, 1 for F0 and a boolean flag for the voiced/unvoiced decision. The prediction model is then trained to minimize the mean squared error loss between prediction and ground truth. This architecture is similar to the acoustic modelling of statistical TTS. We first use a feed-forward DNN as the core of the prediction model, then we use LSTMs~\cite{hochreiter1997long} for better incorporation of context. For the feed-forward DNN, we include an explicit context of $\pm 4$ neighboring frames.

\begin{table*}[bt]
    \centering
    \begin{tabular}{lcc c r@{.}lcc} \toprule
     & \multicolumn{2} {c} {\bfseries Spectral Distortion } & \phantom{abc} & \multicolumn{4} {c} {\bfseries F0 measures} \\ 
     \cmidrule{2-3}  \cmidrule{5-8}
     \multicolumn{1}{c}{System}                     & MCD (dB$\downarrow$) & BAPD (dB$\downarrow$) && \multicolumn{2}{c}{RMSE (Hz$\downarrow$)} & CORR ($\uparrow$) & VUV ($\downarrow$) \\ 
    \midrule
      PR-clean      & 2.68  & 0.16      && 4&95 & 0.96 & 2.78\% \\
      \midrule
    TTS (DNN)                   & 5.28  & 0.25       && 13&06   & 0.71 & 6.66\% \\
    TTS (LSTM)                  & 5.05  & 0.24       && 12&60    & 0.73 & 5.60\% \\
    PR (DNN)  & 5.07 & 0.19     && 8&83 & 0.93 & 6.48\% \\
    PR (LSTM) & \textbf{4.81} & \textbf{0.19}     && \textbf{5}&\textbf{62} & \textbf{0.95} & \textbf{5.27\%} \\
    \bottomrule
    \end{tabular}
    \caption{TTS objective measures for single-speaker experiment: mean cepstral distortion (MCD), band aperiodicity (BAPD), root mean square error (RMSE), voiced-unvoiced error rate (VUV), and correlation (CORR). For MCD, BAPD, RMSE, and VUV lower is better ($\downarrow$), for CORR higher is better ($\uparrow$).}
    \label{tab:obj_eval_tts}
\end{table*}
\section{Experiments}
\label{sec:expr}
\subsection{Dataset}
\label{ssec:data}
The noisy dataset is generated by adding environmental noise to the CMU arctic speech dataset~\cite{kominek2004cmu}. The arctic dataset contains the same 1132 sentences spoken by four different speakers. The speech is recorded in studio environment. The sentences are taken from different texts from Project Gutenberg and are phonetically balanced. We add environmental noise from the CHiME-3 challenge dataset~\cite{BarkerthirdCHiMEspeech2015}. The noise was recorded in four different environments: street, pedestrian walkway, cafe, and bus interior. Six channels are available for each noisy file, we treat each channel as a separate noise recording. We mix clean speech with a randomly chosen noise file starting from a random offset with a constant gain of 0.95. The signal-to-noise ratio (SNR) of the noisy files ranges from $-$6~dB to 21~dB, with average being 6~dB. The sentences are 2 to 13 words long, with a mean length of 9 words. We mainly use a female speech corpus (``slt'') for our experiments. A male (``bdl'') voice is used to test the speaker-dependence of the system. The dataset is partitioned into 1000-66-66 as train-dev-test. 
%
Features are extracted with a window size of 64~ms at a 5~ms hop size. 

\subsection{Evaluation}
\label{ssec:eval}
We evaluate two aspects of the parametric resynthesis system.
Firstly, we compare speech synthesis objective metrics like spectral distortion and errors in F0 prediction with a TTS system. This quantifies the performance of our model in transferring prosody from noisy to clean speech. Secondly, we compare the intelligibility and quality of the speech generated by parametric resynthesis (PR) against two speech enhancement systems, a DNN-predicted ratio mask (DNN-IRM)~\cite{WangTrainingTargetsSupervised2014} and the oracle Wiener mask (OWM)~\cite{ErdoganPhasesensitiverecognitionboostedspeech2015}. The ideal ratio mask DNN is trained with the same data as PR. The OWM uses knowledge of the true speech to compute the Wiener mask and serves as an upper bound on the performance achievable by mask based enhancement systems\footnote{All files are available at \url{http://mr-pc.org/work/icassp19/}}.

A limitation of the proposed method is that the vocoder is not able to perfectly reproduce clean speech, so we encode and decode clean speech with it in order to estimate the loss in intelligibility and quality attributable to the vocoder alone, which we show is small. We call this system vocoder-encoded-decoded (VED). Moreover, we also measure the performance of a DNN that predicts vocoder parameters directly from clean speech as a more realistic upper bound on our speech denoising system. This is the PR model with clean speech as input, referred to as PR-clean.


\begin{table*}[bt]
    \centering
    \begin{tabular}{cccccccr@{.}lcr@{.}l} \toprule
& \multicolumn{2}{c}{\bfseries Speakers} & \phantom{a} & \multicolumn{2}{c}{\bfseries Spectral Distortion} & \phantom{a}& \multicolumn{5}{c}{\bfseries F0 measures}\\
\cmidrule{2-3}  \cmidrule{5-6}  \cmidrule{8-12}
    Model & Train & Test && MCD (dB$\downarrow$) & BAPD(dB$\downarrow$) && \multicolumn{2}{c}{RMSE(Hz$\downarrow$)} & CORR($\uparrow$) & \multicolumn{2}{c}{UUV($\downarrow$)} \\
     \midrule
    PR & slt      & slt    && 4.81 & 0.19 && 5&62  & 0.95 & 5&27\% \\
    PR & slt+bdl & slt    && 4.91 & 0.20 && 8&36  & 0.92 & 6&50\%\\
    PR & bdl      & bdl     && 5.40 & 0.21 && 9&67 & 0.82 & 12&34\%\\
    PR & slt+bdl & bdl    && 5.19 & 0.21 && 10&41  & 0.82 & 12&17\%\\
    \bottomrule
    \end{tabular}
    \caption{TTS objective measures for multi-speaker parametric resynthesis models compared to single speaker model. }
    \label{tab:speaker}
\end{table*}

\subsection{TTS objective measures}
First, we evaluate the TTS objective measures for PR, PR-clean, and the TTS system. We train the feedforward DNN  with 4 layers of 512 neurons each with tanh activation function and the LSTM  with 2 layers of width 512 each. We use adam optimization~\cite{KingmaAdamMethodStochastic2014} and early stopping regularization. For TTS system inputs, we use the ground truth transcript of the noisy speech. As both TTS and PR are predicting acoustic features, we measure errors in the prediction via mel cepstral distortion (MCD), band aperiodicity distortion (BAPD), F0 root mean square error (RMSE), Pearson correlation (CORR) of F0, and classification error in voiced-unvoiced decisions (VUV). The results are reported in Table~\ref{tab:obj_eval_tts}.

Results from PR-clean show that acoustic parameters that generate speech with very low spectral distortion and F0 error can be predicted from clean speech. More importantly, we see from Table~\ref{tab:obj_eval_tts} that PR performs considerably better than the TTS system. It is also interesting to note that the F0 measures, RMSE and Pearson correlation are significantly better in the parametric resynthesis system than TTS. This demonstrates that it is easier to predict acoustic features, including prosody, from noisy speech than from text. We observe that the LSTM performs best and it is used in our subsequent experiments.

\smallskip\noindent\textbf{Evaluating multiple speaker model}
Next we train a PR model with speech from two speakers and test its effectiveness on each speaker's dataset. We first train two single-speaker PR models using the slt (female) and bdl (male) data in the CMU arctic dataset. Then we train a new PR model with speech from both speakers. We measure the objective metrics on both datasets to understand how well a single model can model both speakers. 
These objective metrics are reported in Table~\ref{tab:speaker}, from which we observe that the single-speaker models slightly out-perform the multi-speaker models. On the bdl dataset, however, the multi-speaker model performs better than the single-speaker model in predicting voicing decisions and in MCD. It scores the same in BAPD and F0 correlation, but does worse on F0 RMSE. 
These results show that the same model can be used for multiple speakers.  In future work we will investigate the degree to which a single model can generalize to completely unseen speakers.

\subsection{Speech enhancement objective measures}

We measure objective intelligibility with short-time-objective-intelligibility (STOI)~\cite{taal2010short} and objective quality with perceptual evaluation of speech quality (PESQ)~\cite{rix2001perceptual}. We compare the clean, noisy, VED, TTS, PR-clean speech for reference. The results are reported in Table~\ref{tab:obj_denoise}.

Of the vocoder-based systems, VED shows very high objective quality and intelligibility. This demonstrates that the vocoder is able to produce high fidelity speech when it is fed with acoustic parameters that are exactly correct.  The PR-clean system shows slightly lower intelligibility and quality than VED. The TTS system shows very low quality and intelligibility, but this can be explained by the fact that the objective measures compare the output to the original clean signal. 

For the speech denoising systems, the oracle Wiener mask performs best, because it has access to the clean speech.  While it is an upper bound on mask-based speech enhancement, it does degrade the quality of the speech from the clean by attenuating regions where there is speech present, but the noise is louder.  Parametric resynthesis outperforms the predicted IRM in objective quality and intelligibility.

\begin{table}[bt]
    \centering
\begin{tabular}{ccc}
\toprule
Model & PESQ & STOI \\
\midrule
Clean               & 4.50 & 1.00 \\
VED  & 3.39 & 0.93 \\
OWM      & 3.31 & 0.96 \\
PR-clean  & 2.98 & 0.92\\
\midrule
PR &  2.43 & 0.87 \\
DNN-IRM          & 2.26 & 0.80\\
Noisy               & 1.88 & 0.88\\
TTS                 & 1.33 & 0.08\\
\bottomrule
\end{tabular}
\caption{Speech enhancement objective metrics: quality (PESQ) and intelligibility (STOI), higher is better for both. Systems in the top section use oracle information about the clean speech. All systems sorted by PESQ.}
\label{tab:obj_denoise}
\end{table}

\begin{figure}[tb]
\begin{minipage}[b]{0.95\linewidth}
  \centering
  \centerline{\includegraphics[width=.75\linewidth, trim={0 0.6cm
0 0},clip]{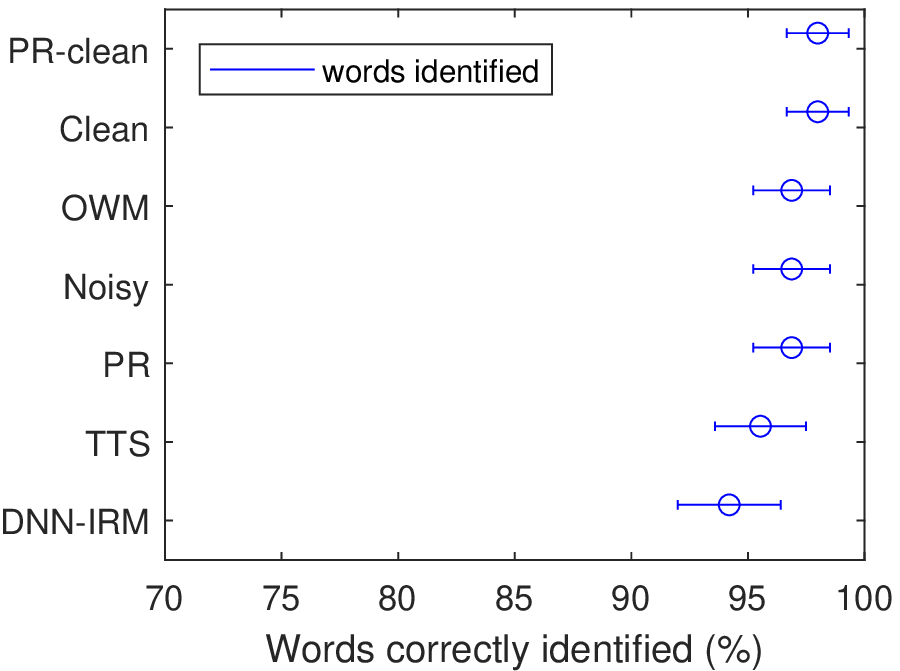}}
  \caption{Subjective intelligibility: percentage of correctly identified words. Error bars show twice the standard error.}
 \label{fig:intel}
\vspace{0.4cm}
\end{minipage}
\begin{minipage}[b]{0.95\linewidth}
  \centering
  \centerline{\includegraphics[width=.8\linewidth,  trim={0 0.61cm
0.4 0},clip]{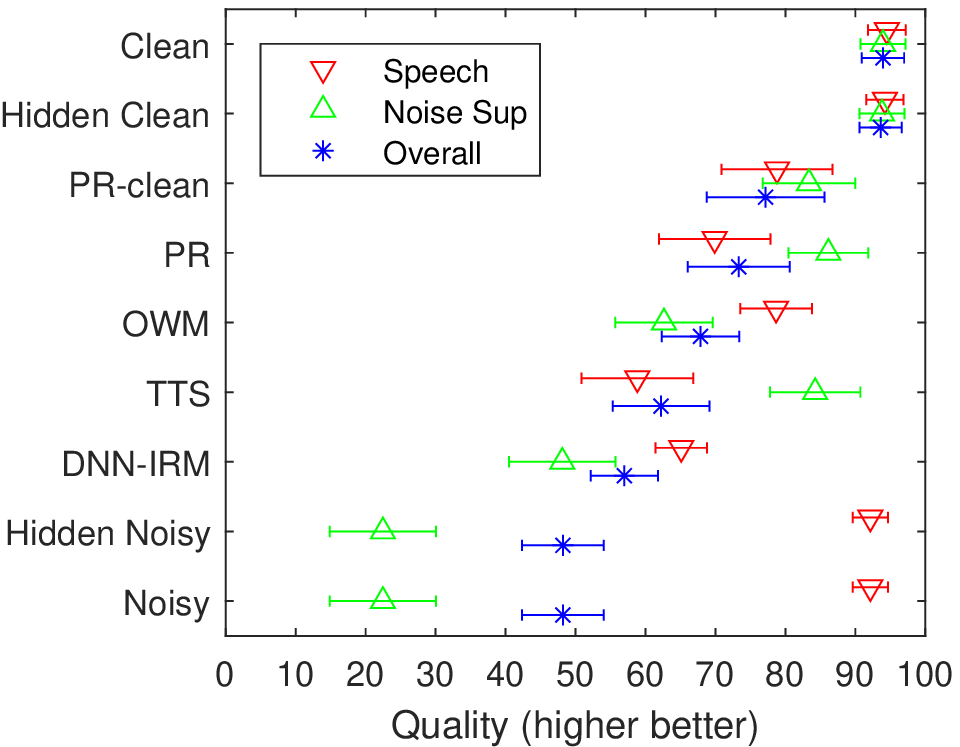}}
  \caption{Subjective quality: higher is better.}
  \label{fig:qual}
\end{minipage}
\end{figure}

\subsection{Subjective Intelligibility and Quality}

Finally we evaluate the subjective intelligibility and quality of PR compared with OWM, DNN-IRM, PR-clean, and the ground truth clean and noisy speech. From 66 test sentences, we chose 12, with 4 sentences from each of three groups: SNR $< 0$ dB, $0$~dB $\leq$ SNR $< 5$~dB, and $5$~dB $\leq$ SNR. Preliminary listening tests showed that the PR-clean files sounded quite similar to the VED files, so we included only PR-clean.  This resulted in a total of 84 files (7 versions of 12 sentences).

For the subjective intelligibility test, subjects were presented with all 84 sentences in a random order and were asked to transcribe the words that they heard in each one.  Four subjects listened to the files. A list of all of the words was given to the subjects in alphabetical order, but they were asked to write what they heard. Figure~\ref{fig:intel} shows the percentage of words correctly identified averaged over all files. Intelligibility is very high ($>90\%$) in all systems, including noisy. PR-clean achieves intelligibility as high as clean speech. OWM, PR, and noisy speech had equivalent intelligibility, slightly below that of clean speech. This shows that PR achieves intelligibility as high as the oracle Wiener mask.

The speech quality test follows the Multiple Stimuli with Hidden Reference and Anchor (MUSHRA) paradigm~\cite{MUSHRA}. Subjects were presented with all seven of the versions of a given sentence together in a random order without identifiers, along with reference clean and noisy versions. The subjects rated the speech quality, noise reduction quality, and overall quality of each version between 1 and 100, with higher scores denoting better quality. Three subjects participated and results are shown in Figure~\ref{fig:qual}.

From the results, we see that the PR system achieves higher noise suppression quality than the OWM, demonstrating that the output is noise-free. PR also achieves comparable overall quality to OWM and PR-clean, indicating that its performance is close to the ceiling imposed by the vocoder.  This ceiling is demonstrated by the difference between PR-clean and the original clean speech.  Note also that the large objective differences between PR and OWM are not present in the subjective results, suggesting that reference-based objective measures may not be accurate for synthetic signals. The PR system achieves better speech quality than the TTS system and better quality in all three measures than DNN-IRM. 

\section{Conclusion}
\label{sec:concl}
This paper has introduced a speech denoising system inspired by statistical text-to-speech synthesis. The proposed  parametric resynthesis system predicts the time-varying acoustic parameters of clean speech directly from noisy speech, and then uses a vocoder to generate the speech waveform. We show that this model outperforms statistical TTS by capturing the prosody of the noisy speech. It provides comparable quality and intelligibility to the oracle Wiener mask by reproducing all parts of the speech signal, even those buried in noise, while still allowing room for improvement as demonstrated by its own oracle upper bound. Future work will explore the extent of speaker-independence that is achievable with this system and other kinds of inputs like filtered and degraded speech~\cite{mandel15d}, and electrophysiological recordings like EEG~\cite{OSullivanAttentionalSelectionCocktail2015} and ECoG~\cite{AkbariReconstructingintelligiblespeech2018}. 

\section{Acknowledgments}

The authors would like to thank Yuxuan Wang for helpful discussions.
This material is based upon work supported by the National Science Foundation (NSF) under Grant IIS-1618061. Any opinions, findings, and conclusions or recommendations expressed in this material are those of the author(s) and do not necessarily reflect
the views of the NSF.


\bibliographystyle{IEEEbib}
\bibliography{refs}

\end{document}